\documentclass[reprint,twocolumn,showpacs,showkeys]{revtex4-1}

\usepackage{latexsym,graphicx,graphics}
\usepackage{appendix}
\usepackage{amsmath,epsfig}

\newcommand{\abs}[1]{\left| #1 \right|}

\newcommand{\bb}[1]{ \mbox{\boldmath$ #1$}}

\newcommand{\Eq}[1]{Eq.~(\ref{#1})}
\newcommand{\Eqs}[2]{Eqs.~(\ref{#1})--(\ref{#2})}

\newcommand{\figstyle}[1]{\small{#1}}

\newcommand{\unit}[1]{\bb{\hat{#1}}}

\begin{document}

\title{Longitudinal chirality, enhanced non-reciprocity, and nano-scale planar one-way plasmonic guiding}
\author{Y. Mazor}
\surname{Mazor}
% \email{hadady@eng.tau.ac.il}
\author{Ben Z. Steinberg}
\surname{Steinberg}
\email{steinber@eng.tau.ac.il}
\thanks{ - Corresponding author. \\This research was supported by the Israel Science Foundation (grant 1503/10)}.
\affiliation{School of Electrical Engineering, Tel Aviv University, Ramat-Aviv, Tel-Aviv 69978  Israel}
\date{February 2012}

\begin{abstract}
 When a linear chain of plasmonic nano-particles is exposed to a transverse DC magnetic field, the chain modes are elliptically polarized, in a single plane parallel to the chain axis; hence, a new chain mode of longitudinal plasmon-rotation is created. If, in addition, the chain geometry possesses longitudinal rotation, e.g.~by using ellipsoidal particles that rotate in the same plane as the plasmon rotation, strong non-reciprocity is created. The structure possesses a new kind of chirality--the longitudinal chirality--and supports one-way guiding. Since all particles rotate in the same plane, the geometry is planar and can be fabricated by printing leaf-like patches on a single plane. Furthermore, the magnetic field is significantly weaker than in previously reported one-way guiding structures. These properties are examined for ideal (lossless) and for lossy chains.
\end{abstract}


\maketitle

\section{Introduction}

Linear chains of identical and equally spaced nano-particles were studied in a number of publications \cite{Quinten}-\cite{EnghetaChain}. They support optical modes with relatively low attenuation and with no radiation to the free space if the inter-particle distance is smaller then the free space wavelength $\lambda$, and then the modes can be much narrower than $\lambda$. Hence the name ``Sub-Diffraction Chains" (SDC). SDCs were proposed as waveguides, junctions, and couplers \cite{Quinten}-\cite{Lomakin2}.

Recently, SDCs were suggested as candidates for one-way guiding \cite{HadadSteinbergPRL}. The physics there is based on creating an interplay of two types of rotations: geometrical chirality and Faraday rotation (caused by longitudinal magnetization). This interplay strongly enhances non-reciprocity and eventually leads to one-way guiding. The attractive features in \cite{HadadSteinbergPRL} are the 3D nano-scale dimensions and the fact that the magnetic field bias is an order of magnitude weaker compared to some alternative structures \cite{YUFAN}.  However, its chiral geometry is difficult to fabricate. Other works on one-way guiding employ periodic structures of non-reciprocal material operating around their Braag point, thus requiring relatively large structures \cite{YUFAN_APL}.

\begin{figure}[htbp]
\vspace*{-0.15in}
    %\centering
    \hspace*{-0.2in}
        \includegraphics[width=7.0cm]{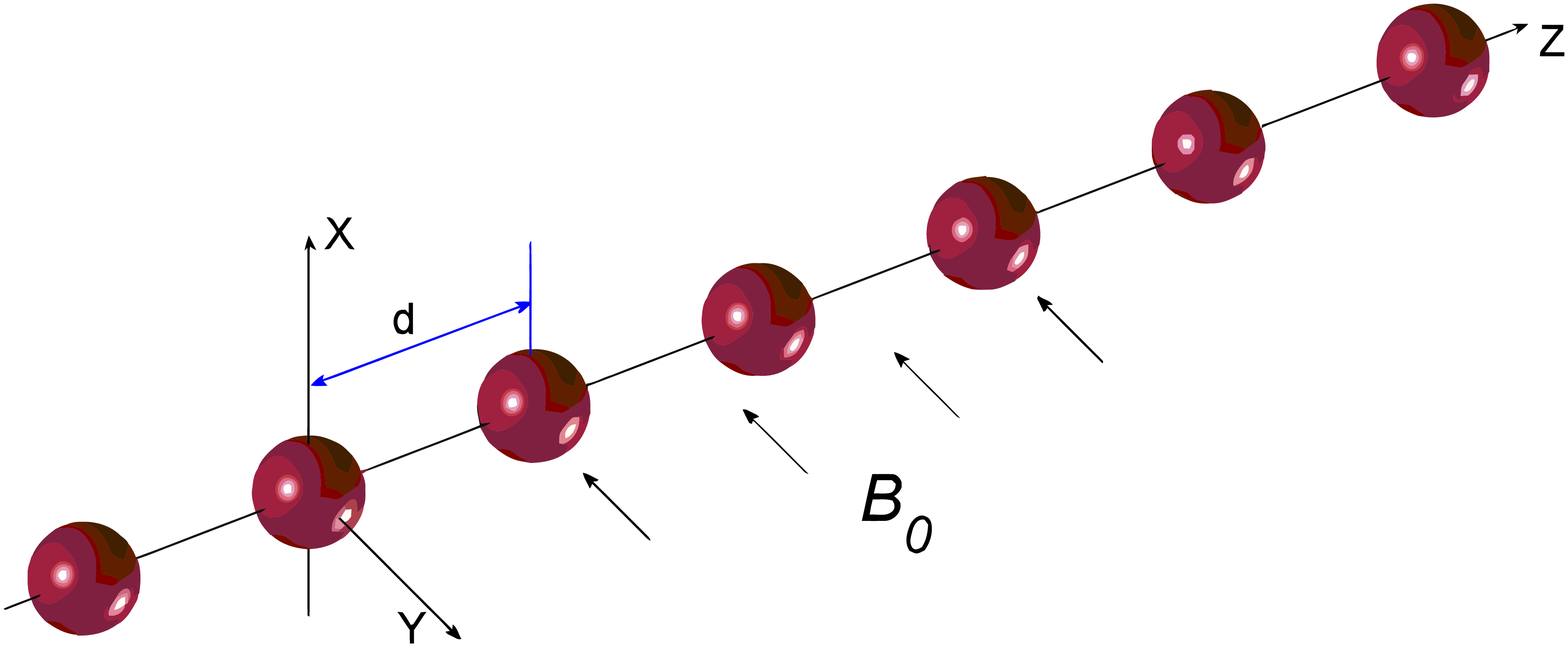}\vspace*{-0.3in}
        \includegraphics[width=7.0cm]{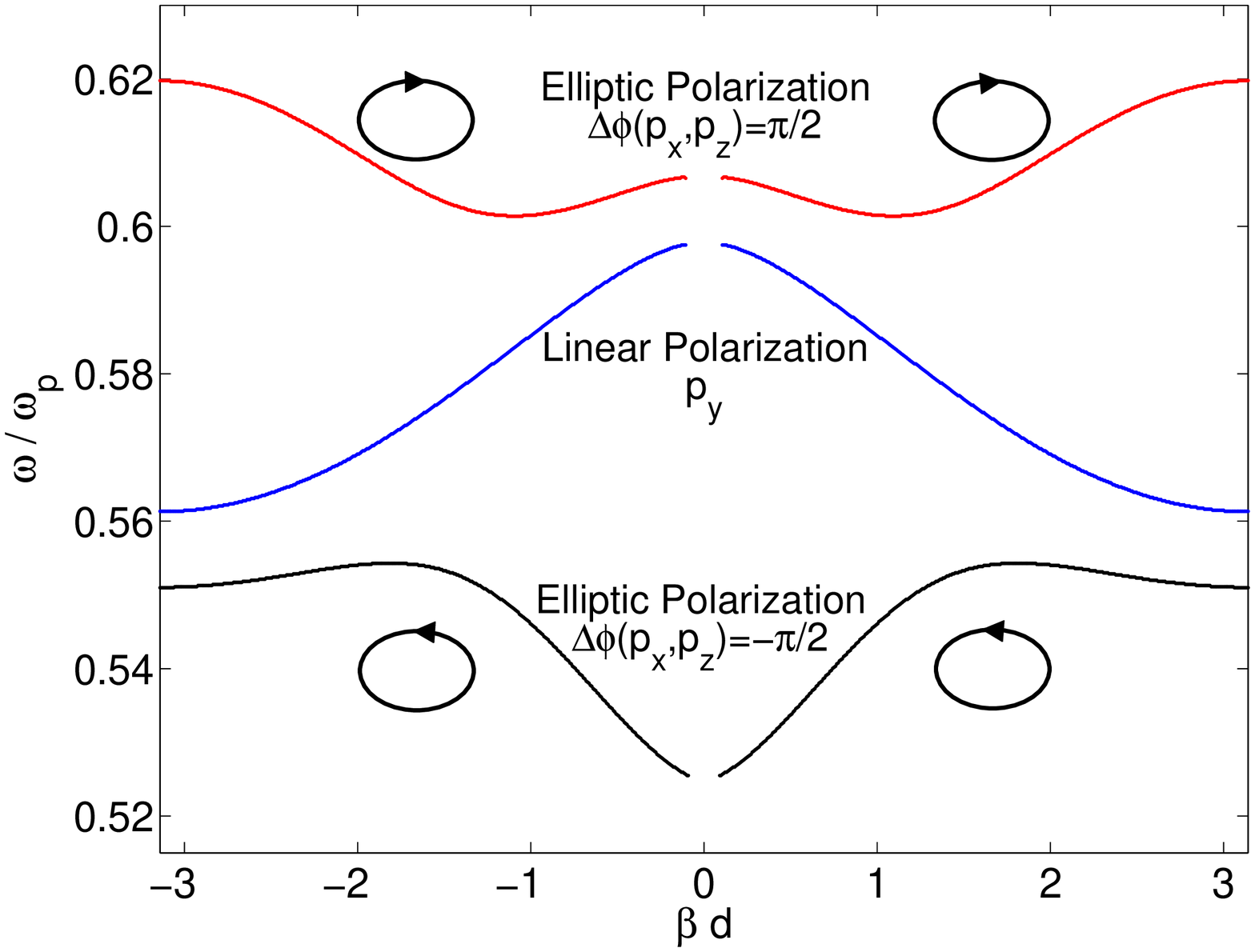}\vspace*{-0.3cm}
    \caption{\figstyle{Top: a chain of spherical plasmonic particles with transverse magnetization. This chain supports modes with \emph{longitudinal} rotation. Bottom: supported modes dispersion.}}
    \label{fig1}
\end{figure}

\begin{figure}[htbp]
\vspace*{-0.15in}
    %\centering
    \hspace*{-0.2in}
        \includegraphics[width=8cm]{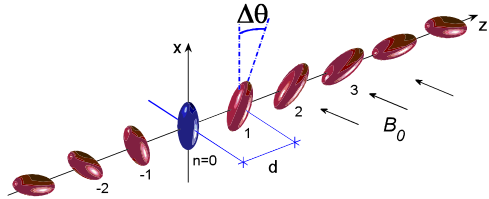}\vspace*{-0.3in}
    \caption{\figstyle{A chain of plasmonic prolate ellipsoids with transverse magnetization and \emph{longitudinal chirality}. This chain supports one-way optical guiding.}}
    \label{fig2}
\end{figure}

 Here we suggest a new type of SDC for one-way optical guiding. The structure is planar in nature, so it can be fabricated by relatively simple printing procedures of leaf-like patches on a single plane. Moreover, it requires magnetization field that is significantly weaker than in \cite{HadadSteinbergPRL, YUFAN}. The underlying physics and geometry are described in Figs.~\ref{fig1}-\ref{fig2}. A conventional SDC of spherical plasmonic particles supports three independent electric dipole modes; one with longitudinal polarization $p_z$, and two degenerate modes with transverse and mutually orthogonal polarizations $p_x,p_y$ \cite{EnghetaChain}. If this SDC is exposed to transverse magnetization, as shown in Fig.~\ref{fig1}, the $p_y$ mode is unaffected, but $\bb{B}_0=\unit{y}B_0$ couples $p_x$ and $p_z$. This coupling creates two new modes of elliptical polarizations in the $x,z$ plane, with two new dispersion curves (see analysis below). At any operation frequency, one wave propagates in the direction of rotation (``paddles'' forward and ``rides'' forward) and the second propagates counter to the rotation (paddles forward and rides backward). Hence, these new SDC modes are non-reciprocal. However, their dispersion is still reciprocal (i.e.~even in $\beta$). The route to enhanced non-reciprocity and one-way guiding is to create an interplay of \emph{two-type rotations}. Hence, we add a \emph{longitudinal chirality}: we replace the spheres by ellipsoids, rotated in the same plane of the elliptical polarization, i.e.~in the $x,z$ plane as shown in Fig.~\ref{fig2}. The rotation step is $\Delta\theta$. As we show below, this chain indeed supports one-way guiding. It is periodic only for rational $\Delta\theta/\pi$. If $\Delta\theta/\pi=n/m$ and $m,n$ are coprime, then the period consists of $m$ particles. Unlike the spiral structure in \cite{HadadSteinbergPRL}, if this ratio is irrational then \emph{there is no coordinate transformation under which the chain becomes periodic.}

Below we use the Discrete Dipole Approximation (DDA) and polarizability theory. These are standard tools in SDCs analysis \cite{Brongersma}-\cite{Lomakin2}. They hold when the particle radius $a$ and inter-particle distance $d$ satisfy $a\ll\lambda$ and $a\ll d$, but studies show good accuracy even for $d=3a$ \cite{MaierKikAtwater}.

\section{Formulation}

If a small particle with polarizability $\bb{\alpha}$ is subject to an electric field  whose local value in the \emph{absence of the particle} is $\bb{E}^L$, its response is described by the electric dipole $\bb{p}=\bb{\alpha}\bb{E}^L$.
The dynamic tensor-polarizability of a general ellipsoid whose principal axes are aligned with the $x,y,z$ axes, made of an anisotropic material with electric susceptibility $\bb{\chi}$, is obtained via
\begin{equation}
\epsilon_0\bb{\alpha}^{-1}=V^{-1}(\bb{\chi}^{-1}+{\bf L})-\frac{ik^3}{6\pi}{\bf I}.
\label{eq1}
\end{equation}
Here $k$ is the vacuum wavenumber, ${\bf I}$ is the $3\times 3$ identity matrix, and the imaginary term $ik^3(6\pi)^{-1}{\bf I}$ represents radiation loss. $V=4\pi a_xa_ya_z/3$ is the ellipsoid volume with $a_x,a_y,a_z$ its principal semi-axes, and
 ${\bf L} =\mbox{diag}(N_x,N_y,N_z)$ is the depolarization matrix whose entries are obtained from $a_x,a_y,a_z$ by elliptic integrals and satisfy $\sum_uN_u=1$ \cite{SihvolaBook}. Under the Drude model and magnetization $\bb{B}_0=\unit{y}B_0$, $\bb{\chi}$ is \cite{JACKSON}
\begin{equation}
\bb{\chi}=\frac{-\bar{\omega}^{-2}}{(\bar{\omega}+i\sigma)^2-\bar{\omega}_b^2}\!
 \left(
\begin{array}{ccc}
\bar{\chi}_{xx} & 0 & \bar{\chi}_{xz}\\
0 & \bar{\chi}_{yy} & 0\\
-\bar{\chi}_{xz} & 0& \bar{\chi}_{zz}\end{array}\right),\label{eq2}
\end{equation}
with $\bar{\chi}_{xx}=\bar{\chi}_{zz}= \bar{\omega}^2+i\sigma\bar{\omega}$,
$\bar{\chi}_{yy}= \bar{\chi}_{xx}-\bar{\omega}\bar{\omega}_b^2/(\bar{\omega}+i\sigma)$, and
$\bar{\chi}_{xz}= i\bar{\omega}\bar{\omega}_b$, and
where $\bar{\omega}=\omega/\omega_p$, $\bar{\omega}_b=\omega_b/\omega_p$, and where $\omega_p$ and $\omega_b=-q_eB_0/m_e$ are the plasma and cyclotron frequencies. $\sigma=(\tau\omega_p)^{-1}$ represents loss, where $\tau$ is the dissipation time constant.
Equations (\ref{eq1})-(\ref{eq2}) fully describe the ellipsoid at the origin.
The $n$-th particle polarizability, $\bb{\alpha}_n$, is given by the $n\Delta\theta$ \emph{longitudinal rotation} (about $\unit{y}$)
\begin{equation}
\bb{\alpha}_n={\bf T}_{-n}^\ell\bb{\alpha}{\bf T}_n^\ell, \label{eq3}
\end{equation}
where the non-zero entries of ${\bf T}_n^\ell$ are $t_{11}=t_{33}=\cos n\Delta\theta$, $t_{22}=1$, $t_{13}=-t_{31}=\sin n\Delta\theta$.
The electric field at $(0,0,z)$ due to a short dipole $\bb{p}$ at $(0,0,z')$ is given by the matrix relation $\bb{E}(z)=\epsilon_0^{-1}{\bf A}(z-z') \,\bb{p}$ where
\begin{equation}
{\bf A}(z)=\frac{e^{ik\abs{z}}}{4\pi\abs{z}}\,\left[k^2{\bf A}_1 + \left(\frac{1}{z^2}-\frac{ik}{\abs{z}}\right){\bf A}_2\right]
\label{eq4}
\end{equation}
here ${\bf A}_1=\mbox{diag}(1,1,0),\,{\bf A}_2=\mbox{diag}(-1,-1,2)$. We express now the local exciting field of the $m$-th particle in the chain as a sum of contributions from all its neighbor-dipoles, and apply $\bb{\alpha}_m$. The result relates the $m$-th dipole excitation $\bb{p}_m$ to its neighbors
\begin{equation}
\bb{p}_m=\epsilon_0^{-1}\bb{\alpha}_m\!\! \sum_{n,\,n\ne m}\!\! {\bf A}[(m-n)d]\, \bb{p}_n. \label{eq5}
\end{equation}
This equation is not shift-invariant. Furthermore, since ${\bf T}_n^\ell$ is a rotation about $\unit{y}$ and not about the longitudinal axis $\unit{z}$, it does not commute with the propagator ${\bf A}$ so the mathematical transformation used in \cite{HadadSteinbergPRL} cannot be applied. However, we set $\Delta\theta/\pi=N/M$ rational, hence the period $D=Md$ consists of $M$ particles with polarizabilities $\bb{\alpha}_0,\bb{\alpha}_1,\ldots\bb{\alpha}_{M-1}$ as in \Eq{eq3} and
$\bb{\alpha}_{m+lM}=\bb{\alpha}_m\,\forall$ integer $l$. Then by periodicity
\begin{equation}
\bb{p}_{n+lM}=\bb{p}_{n}e^{i\beta lMd},\label{eq7}
\end{equation}
and the chain modes are determined from the $M$ vectors of the reference period $\bb{p}_0,\ldots,\bb{p}_{M-1}$. We substitute this solution into \Eq{eq5}. For each $m$ within a period we decompose the infinite sum into a sum $\bb{S}_0$ of contributions from identical particles ($n=m+lM,\,l\ne 0$), and a set of summations $\bb{S}_{m-n'}$ of contributions from the rest ($n=n'+lM,\, n'\ne m$). The result is the $3M\times 3M$ matrix equation
\begin{equation}
(\epsilon_0\bb{\alpha}_m^{-1}-\bb{S}_0)\bb{p}_m
-\sum_{n=0,n\ne m}^{M-1}\bb{S}_{m-n}\bb{p}_n=0,
\label{eq8}
\end{equation}
with $m=0,\ldots,M-1$, and where
\begin{equation}
\bb{S}_q\! =\!\left\{\!
\begin{array}{ll}
\sum_{l\ne 0}\!\bb{A}(lD)e^{i\beta l D}, & q=0,\\
&\\
\sum_l\!\bb{A}(qd-lD)e^{i\beta l D}, & 1\!\le\!\abs{q}\! <M\! .\\
\end{array}\right.
\label{eq9}
\end{equation}
The modes are obtained by looking for $\omega(\beta)$ at which \Eq{eq8} determinant vanishes. The vectors that span the corresponding nullspace are $\bb{p}_0,\ldots,\bb{p}_{M-1}$ that with \Eq{eq7} describe the entire chain excitation.

Note that the series for $\bb{S}_q$ above converge poorly. However, they can be cast in terms of the Polylogarithm functions $Li_s$, for which efficient summation formulas exist (see \cite{EnghetaChain,PolyLogarithmBook} and Appendix). First observe that
the matrix $\bb{S}_0$ is identical to that obtained in conventional chains with $d=D$. Hence \cite{EnghetaChain}
\begin{equation}
\bb{S}_0= \frac{k^3}{4\pi}\sum_{s=1}^3 u_s f_s(kD,\beta D)\bb{A}_s,
\label{eq10}
\end{equation}
where $(u_1,u_2,u_3)=(1,-i,1), \bb{A}_3=\bb{A}_2$ and
\begin{equation}
f_s(x,y)=x^{-s} [Li_s(e^{ix+iy})+Li_s(e^{ix-iy})]\label{eq11}
\end{equation}
where $Li_s(z)\equiv\sum_{n=1}^\infty \frac{z^n}{n^s}$ is the $s$-th order Polylogarithm function. To write $\bb{S}_q$ in terms of $Li_s$, note that all sums in $\bb{S}_q$ have the general form
\begin{equation}
\zeta=\sum_{l=0}^\infty\frac{(e^{i\xi})^{lM+q}}{(lM+q)^s}=\sum_{n=1}^\infty\frac{(e^{i\xi})^n}{n^s}\cdot a_n(q),\label{eq12}
\end{equation}
where $a_n(q)=a_{n+M}(q)$ is $M$-periodic sequence with $a_n(q)=0$ for $1\le n\ne q\le M$, and $a_q(q)=1$. But $Ma_n(q)=\sum_{r=0}^{M-1}e^{i2\pi r(n-q)/M}$, hence
\begin{equation}
\zeta=M^{-1}\sum_{r=0}^{M-1}e^{-i2\pi r q/M}Li_s(e^{i\xi+i2\pi r/M}).\label{eq13}
\end{equation}
Following the standard steps leading from \Eq{eq9} to \Eq{eq10} and using \Eq{eq13}, we get for $\bb{S}_q,\,1\le\abs{q}
<M$ the same expression as in \Eq{eq10} but with $f_s$ replaced by $h_s(kd,\beta d;q)$,
\begin{equation}
h_s(x,y;q)=\frac{e^{iyq}}{M}
\sum_{r=0}^{M-1} e^{-i2\pi\frac{rq}{M}} f_s\left(x,y-\frac{2\pi r}{M}\right).\label{eq14}
\end{equation}

\section{Examples}

We turn to study the structure. First, we examine a transversely magnetized chain of \emph{spherical} particles of radius $a$. It is a special case of \Eq{eq8} with the sum over $\bb{S}_q,q\ne 0$ dropped, $M=1 (D=d)$, $\Delta\theta=0$, and $\bb{\alpha}_0=\bb{\alpha}$ given by \Eqs{eq1}{eq2} with all depolarization factors set to $1/3$. The chain dispersion is obtained by $\mbox{Det} (\epsilon_0\bb{\alpha}^{-1}-\bb{S}_0)=0$. For lossless chain, this can be satisfied only if $\mbox{Im}[\mbox{diag}(\bb{S}_0)]=-k^3/(6\pi)$. Also, the first and third rows of \Eq{eq8} now are linearly dependent, and read
\begin{equation}
\left\{\bar{\omega}^2\! -\! 1/3 \! +\!  V[(\bb{S}_0)_{11}\! +\! ik^3/(6\pi)]\right\}p_x
= i\bar{\omega}\bar{\omega}_bp_z.\label{eq15}
\end{equation}
 However, as discussed above the term in the square brackets must be real. Hence there is a phase difference of $\pm\pi/2$ between $p_z$ and $p_x$ if $\bar{\omega}_b\ne 0$. This implies that the chain modes are elliptically polarized in the $(x,z)$ plane.

The chain dispersion for $d=3a=\lambda_p/30$ and no loss was calculated numerically, and is shown in Fig.~\ref{fig1}. We have applied a relatively strong magnetization of $\omega_b=0.05\omega_p$, in order to observe the features clearly (the one-way property shown below is obtained at much weaker magnetizations). The $\unit{y}$-polarized mode is identical to that of a conventional chain. Two additional modes of elliptic polarization in the $x-z$ plane exist in the chain. When observed from $y>0$, one rotates clockwise [shown by the upper (red) curve] and the second rotates counter-clockwise [shown by the lower (black) curve]. This elliptic polarization is only due to the transverse magnetization. It is mathematically evident from the $\pm\pi/2$ phase difference between the $\unit{x}$ and $\unit{z}$ components we have observed in the solutions for $\bb{p}_0$, as seen in \Eq{eq15}. All dispersion curves are even in $\beta$, permitting propagation in both $+z$ and $-z$ directions. Also all dispersion curves possess the light-line gap clearly seen in the center. However, the light-line dispersion branches that run parallel to the light-line cone $\beta=\omega/c$ and that are associated with transverse polarization in conventional chains \cite{EnghetaChain, HadadSteinberg_PRB}, represent modes that are practically not excited \cite{HadadSteinberg_PRB} in the elliptically polarized curves (red and black). To avoid cluttering the figure they are not shown here.

\begin{figure}[htbp]
\vspace*{-0.0in}
    \centering
    \hspace*{-0.25in}
        \includegraphics[width=8.5cm]{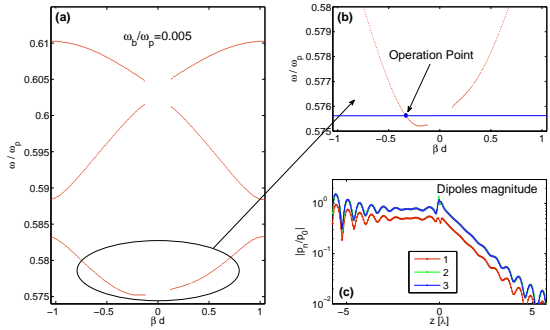}\vspace*{-0.2in}
    \caption{\figstyle{One-way guiding in the upper band. (a) Dispersion curves. The gap in Fig.\ref{fig1} is shifted rightward by the longitudinal chirality. (b) Frequency selection for one-way guiding. (c) Chain response.}}
    \label{fig3}
\end{figure}

\begin{figure}[htbp]
\vspace*{-0.0in}
    \centering
    \hspace*{-0.25in}
        \includegraphics[width=8.5cm]{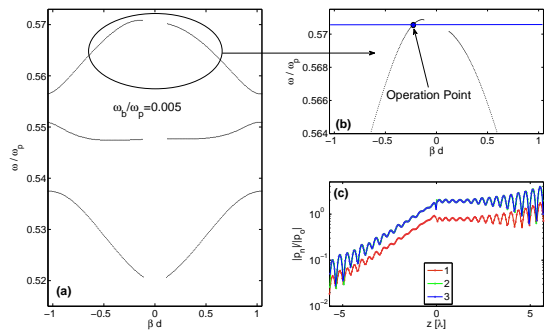}\vspace*{-0.2in}
    \caption{\figstyle{One-way guiding in the lower band.}}
    \label{fig4}
\end{figure}

Next, we add \emph{longitudinal chirality};
we replace the spheres by prolate ellipsoids with semi-axis $a_x=\lambda_p/90$, axes ratio of $a_y=a_z=0.9a_x$ (nearly spheres), and add longitudinal chirality with $\Delta\theta=\pi/3$ (hence chain period is $D=3d=\lambda_p/10$). Figure \ref{fig3} shows the results for the relatively weak magnetization of $\omega_b=0.005\omega_p$. Due to the added chirality, the upper (red) dispersion curve of Fig.~\ref{fig1} that corresponds to elliptical polarization with clockwise longitudinal rotation, splits to three branches shown in Fig.~\ref{fig3}a, and shifts the light-line gap rightward. Hence symmetry is broken and one way guiding is supported at frequencies within the shifted gap, as shown in Fig.~\ref{fig3}b. Finally, in Fig.~\ref{fig3}c we show the normalized response $\abs{\bb{p}_n}/\abs{\bb{p}_0}$ of a finite chain of $N=600$ particles with the parameters above, for a unit amplitude excitation of the central particle (at the origin), at the selected frequency. This response is obtained by solving \Eq{eq5} for a finite number of particles, with only the central particle forced to oscillate at the aforementioned frequency (hence it is a matrix equation of $3(N-1)$ unknowns). The three curves in Fig.~\ref{fig3}c correspond to the three particles within each period. One way guiding is clearly observed (note the logarithmic scale).
Essentially the same picture holds for the lower (black) dispersion curve of Fig.~\ref{fig1} that corresponds to elliptical polarization with counter-clockwise longitudinal rotation, when the above longitudinal chirality is added to the structure. This is shown in Fig.~\ref{fig4}.

We note that these results are obtained with a very slight breach of spherical symmetry ($a_z=0.9a_x$). This low ellipticity is sufficient to create one-way guiding if $\bb{B}_0=\unit{y}B_0$ is present. In addition, $B_0$ is considerably weaker than that required in previous studies on one-way plasmonic waveguides.

Next, we study chains with loss. Since a \emph{real} dispersion $\omega(\beta)$ does not exist, we have solved \Eq{eq5} for a finite chain with $N=250$ and for the same geometrical parameters ($\Delta\theta, d$) as in the lossless example of Fig.~\ref{fig4}.
\begin{figure}[htbp]
\vspace*{-0.0in}
    \centering
    \hspace*{-0.25in}
        \includegraphics[width=8.0cm]{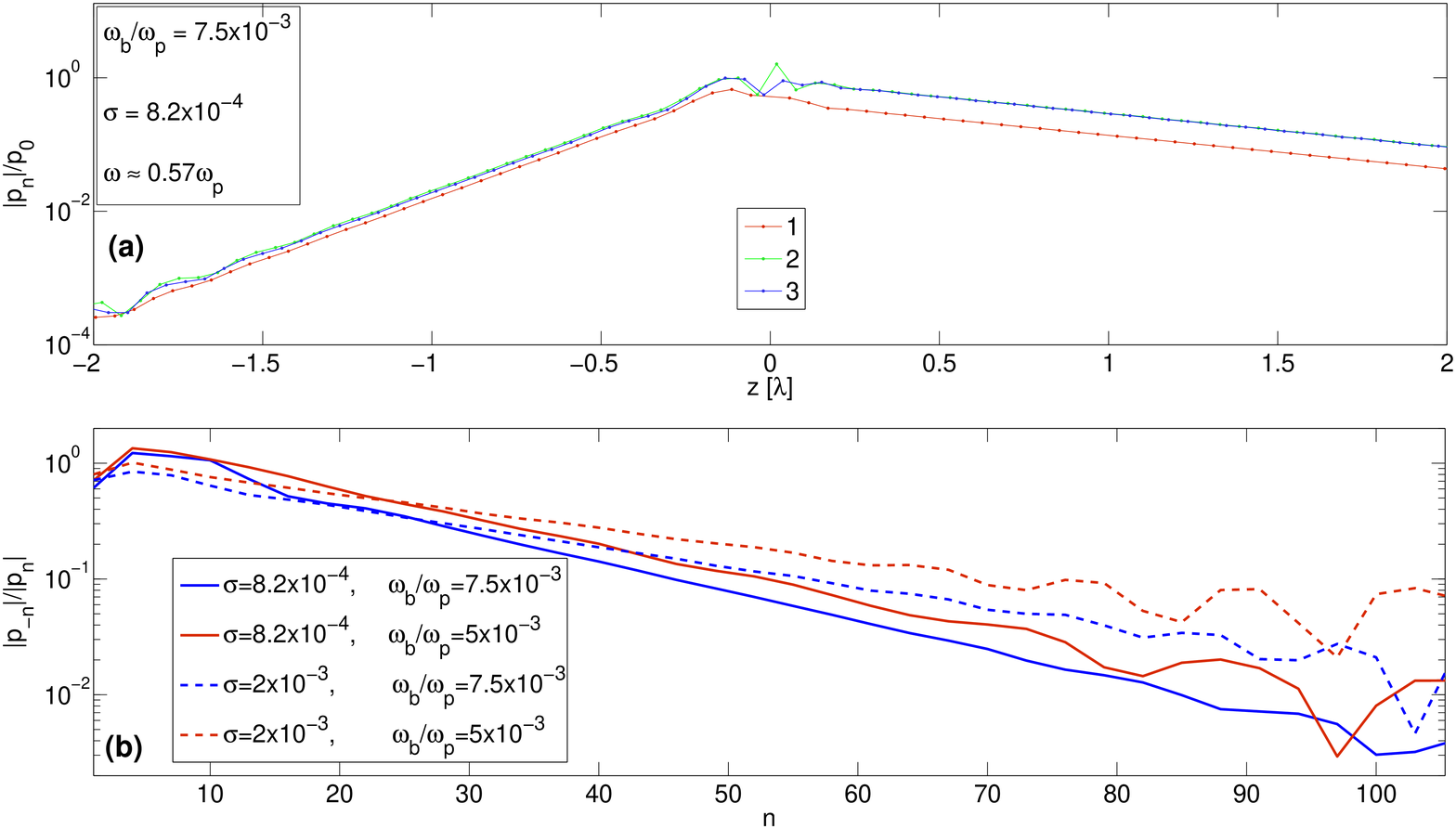}\vspace*{-0.2in}
    \caption{\figstyle{One-way guiding in a lossy chain. (a) Chain response. (b) Isolation ratio.}}
    \label{fig5}
    \end{figure}
The operation frequency $\omega$ and axes ratio $r=a_z/a_x$ were chosen by scanning over the neighborhoods of $\omega, r$ for which the corresponding lossless chain possesses one-way property. The results are shown in Fig.~\ref{fig5}. Figure \ref{fig5}a shows the chain response for loss parameter $\sigma=(\tau\omega_p)^{-1}=8.2\times 10^{-4}$ corresponding to Palladium (Pd) \cite{OptProp}, and magnetization of $\omega_b=7.5\times 10^{-3}\omega_p$. It is obtained with $\omega\approx 0.57\omega_p$ and $r=0.92$. For a clear assessment of the one-way property as a function of magnetization and loss, we plot in Fig.~\ref{fig5}b the ``isolation ratio'' $|\bb{p}_{-n}|/|\bb{p}_{n}|$ vs.~$n$ for various values of loss and magnetization. Again, one-way guiding is evident.

\begin{figure}[htbp]
\vspace*{-0.25in}
    \centering
    \hspace*{-0.25in}
        \includegraphics[width=8cm]{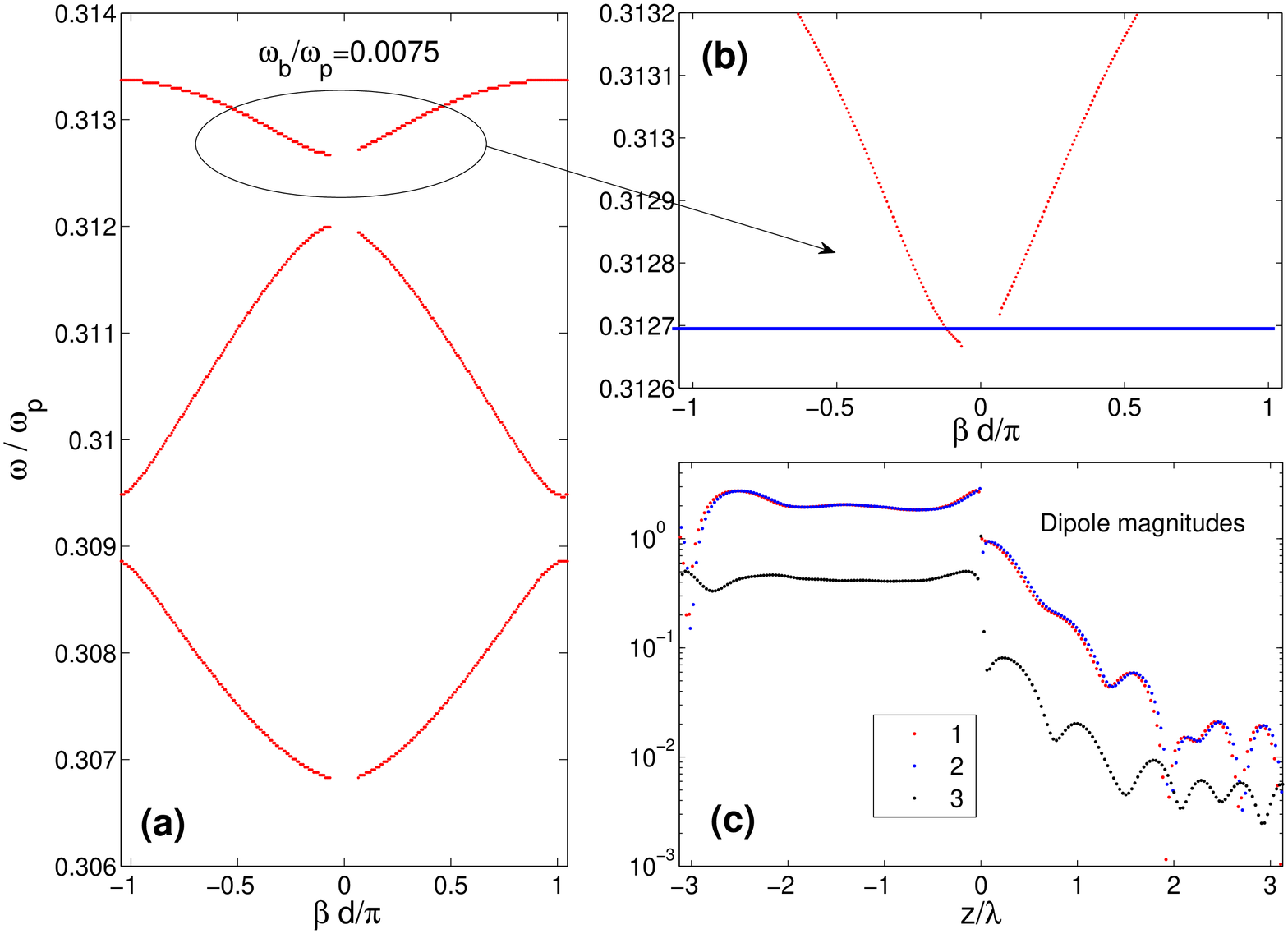}\vspace*{-0.2in}
    \caption{\figstyle{The same as Fig.\ref{fig3} but for flat geometry with $a_x:a_y:a_z=1:0.1:0.75$.}}
    \label{fig6}
\end{figure}

In the examples above the particles are nearly spheres. These examples are important from the theoretical/physical point of view. They show that by using our approach of two-type rotation interplay a very slight breach of symmetry is sufficient to create a profound non-reciprocity and one-way guiding. However from the practical point of view structure \emph{flatness} is at least as important, since it opens the way to employ planar fabrication technologies. To demonstrate this option we have simulated a chain with the following properties. Axes ratio is $a_x:a_y:a_z=1:0.1:0.75$, hence the particle's size in the $\unit{y}$-direction is an order of magnitude smaller then its dimensions in the $x,z$ plane. Since the longitudinal chirality is obtained by rotating the particles about their $y$-axis, it follows that the entire structure is practically flat and lies in the $x,z$ plane. Other geometrical parameters are as before, with magnetization level of $\omega_b=0.0075\omega_p$. A strong one-way property exists and it is shown in Fig.~\ref{fig6}.

\begin{figure}[htbp]
\vspace*{-0.0in}
    \centering
    \hspace*{-0.0in}
        \includegraphics[width=7cm]{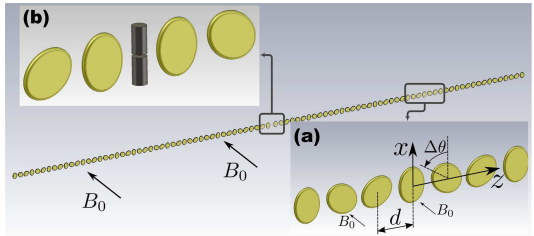}\vspace*{-0.2in}
    \caption{\figstyle{A flat chain geometry for one-way guiding. (a) Geometry details. (b) Excitation by a short dipole located in between two identical chains.}}
    \label{fig7}
\end{figure}

\begin{figure}[htbp]
\vspace*{-0.0in}
    \centering
    \hspace*{-0.25in}
        \includegraphics[width=8.cm]{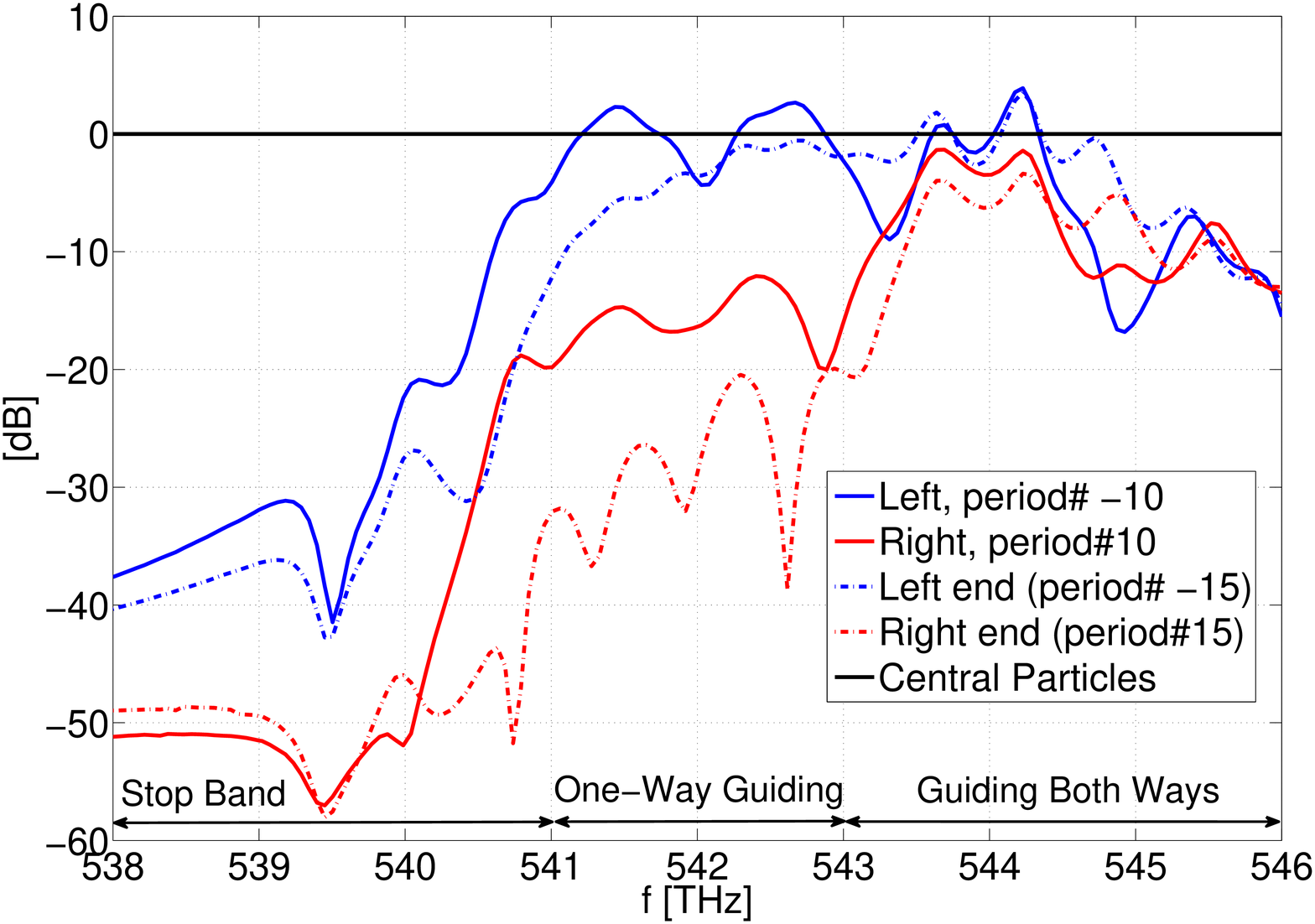}\vspace*{-0.2in}
    \caption{\figstyle{E-fields just above the upright particle in periods\#$\pm 10$ and $\pm 15$, obtained by full-wave solution of the chain shown in Fig.~\ref{fig7}. For each frequency, the values are normalized to the field above the upright particle adjacent to the dipole from left, shown in inset (b) of Fig.~\ref{fig7}.}}
    \label{fig8}
\end{figure}

\begin{figure}[htbp]
\vspace*{-0.1in}
    \centering
    \hspace*{-0.25in}
        \includegraphics[width=8.cm]{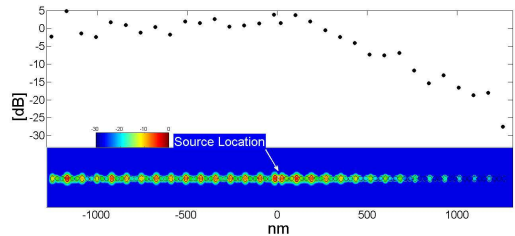}\vspace*{-0.35in}
    \caption{\figstyle{Full-wave solution for the E-field in the $y=0$ plane, at 542THz. Normalized fields in the center of each upright particle are shown on top.}}
    \label{fig9}
\end{figure}

Finally, we verify our approach via a full-wave numerical solutions using the CST software package \cite{CST} for the flat structure shown in Fig.~\ref{fig7}.
The particles are disks of $2.5$nm thickness and of elliptical shape in the $x,z$ plane with long and short diameters of $25$nm and $20$nm, respectively. The inter-particle distance is $d=27.5$nm. The longitudinal chirality angle is $\Delta\theta=-60^\circ$. The particles material plasma frequency $\omega_p$ and loss parameter $\tau^{-1}$ are $2\pi\times 2000$THz and $2$THz, respectively (corresponding to Ag). Due to a limited availability of computing power and memory resources, we can apply the full-wave solution to chains of about 100 particles or less. Hence our chain consists of 92 particles that create two \emph{identical} chains of 46 particles each, one chain on each side of a short dipole antenna that serves as a local source - see inset (b) in Fig.~\ref{fig7}.
Since this geometry should show significant one-way guiding within a distance of 46 particles only (instead of within hundreds of particles used to obtain the results of Fig.~\ref{fig5}) we need to apply stronger magnetization. Hence we use $\omega_b=0.03\omega_p$. We have used the DDA for a first estimate of the chain operation parameters, and then used the CST to search more accurately around the initial guess and to get full-wave solutions.
The full-wave based chain response as a function of frequency is shown in Fig.~\ref{fig8}.
One-way behavior can be clearly observed within the frequency band of 541-543 THz. The ratio between the chain excitation on left and right side of the chain--the isolation ratio--is in the order of 20dB.
Finally, Fig.~\ref{fig9} shows the E-field in the $y=0$ plane at the central frequency of the one-way band, 542THz.

Note that in this frequency $\lambda\approx0.55\mu m$. Hence a significant one way behavior is clearly observed over distances of $O(\lambda)$.

\section{Conclusion}

When an interplay of two-type rotations is supported in a guiding structure, strong non-reciprocity and one-way guiding are created \cite{HadadSteinbergPRL}. In this work, a new geometry possessing longitudinal chirality in sub-diffraction particle chains is proposed. When combined with transverse magnetization, the resulting elliptical rotation of the chain dipole-modes and the geometrical rotation (longitudinal chirality) provide the two-type rotation that lead to strong non-reciprocity and one-way guiding. The required magnetization is considerably weaker than in previously reported studies. This one-way chain can be fabricated by thin-layer printing of leaf-like flakes of thin metals. Since single-atom layers of Graphene may behave as thin metal flakes with controllable properties, exhibiting SPP like behavior with very low loss \cite{GrapheneEngheta}, the present work may offer a basic scheme for one-way guiding on Graphene layer. This is a subject of ongoing research.

\appendix

\section{Evaluation of $Li_s(e^{i\theta})$}

$Li_s(e^{ix})$ for real $x$, can be expressed as finite sums of Clausen integrals \cite{PolyLogarithmBook},\cite{Stegun}. The latter can be computed using series that converge much faster than the algebraic convergence of the formal definition in \Eq{eq9}. Note that $Li_1(z)=-\ln (1-z)$. Furthermore, from \Eq{eq9} it follows that
\begin{equation}
Li_{s+1}(z)=\int_0^z t^{-1}Li_s(t)\, dt\label{eqAp1}
\end{equation}
Collecting the Clausen integrals components that constitute our $Li_s$ and rearranging, we obtain for $s=2$ and $\abs{x}\le\pi$
\begin{eqnarray}
Li_2(e^{ix})&=& \frac{\pi^2}{6}-\frac{x}{4}\left(2\pi-x\right)\label{eqAp2}\\
-&i& \left[x\ln x - x -\frac{1}{2}\sum_{n=1}^\infty \frac{B_{2n}x^{2n+1}}{n(2n+1)(2n)!}\right]\nonumber
\end{eqnarray}
where $B_{2n}$ are the Bernoulli numbers absolute value. With terms up to $n=5$ in the series above, the relative error for $x=\pi$ is in the order of $10^{-5}$. From the integral relation \Eq{eqAp1} we obtain
\begin{eqnarray}
Li_3(e^{ix})&=& \zeta(3)+\frac{x^2}{2}(\ln x-3/2)\nonumber\\
 &-&\frac{1}{4}\sum_{n=1}^\infty\frac{B_{2n}x^{2n+2}}{n(n+1)(2n+1)(2n)!}\nonumber\\
 &+& i\left(\frac{\pi^2 x}{6}-\frac{\pi x^2}{4}+\frac{x^3}{12}\right)\label{eqAp3}
 \end{eqnarray}
 where $\zeta(3)\approx 1.2020569$. Higher order Polylogarithm functions are not needed in the present work.

\end{document}